\documentclass{elsart}
\usepackage{graphicx}
\usepackage{amsmath}
\usepackage{amssymb}
\usepackage{mathrsfs}

\begin{document}

\begin{frontmatter}

\title{
Violation of Bell's inequality in electronic Mach-Zehnder interferometers
}

\author{Kicheon Kang\thanksref{thank1}} and
\author{Kahng Ho Lee}

\address{Department of Physics, Chonnam National University,
 Gwangju 500-757, Korea}

\thanks[thank1]{Corresponding author. E-mail: kicheon.kang@gmail.com}

\begin{abstract}
We propose a possible setup of testing the Bell's inequality
in mesoscopic conductors. The particular implementation uses two coupled
electronic Mach-Zehnder interferometers in which electrons are injected into
the conductors in the quantum Hall regime.
It is shown that the Bell's inequality is violated for an arbitrary coupling
strength between the two interferometers.
\end{abstract}

\end{frontmatter}


One of the most striking features of quantum theory is the
entanglement that gives rise to strong nonlocal correlations between
spatially separated particles. The peculiar properties of
entanglement are inconsistent with any local realistic theories and
can be tested through Bell's inequality (BI)~\cite{bell64}.
%
Experiments with entangled photon pairs have shown violation of the
BI~\cite{bi-optics}. However, no violation of BI with electrons
has been realized yet. Recent advance in nanofabrication
technology makes it possible to study fundamental problems of quantum theory
such as entanglement and measurement in mesoscopic devices.

In this paper, we propose a possible realization of the violation of
BI in mesoscopic conductors. The schematic diagram of Fig.~1
represents two coupled electronic Mach-Zehnder
interferometers~\cite{ji03} in which electrons are injected into the
conductors in the quantum Hall regime. This setup has been
considered for studying complementarity principle that enables a
choice of wavelike or particlelike behavior of
electrons~\cite{kang07}. A recent experiment has reported which-path
(WP) detection and recovery of interference in correlation
measurement in an electronic Mach-Zehnder interferometer by using
the quantum Hall edge channels~\cite{neder07}, which demonstrates
that our scheme can also be implemented with current technology.
%

The two interferometers are electrostatically coupled to each other so that
the Coulomb interaction in the contact region gives rise to modification
of the trajectory leading to a phase shift (denoted by $\Delta\phi$).
No electron exchange is allowed between the two interferometers.
The phase shift
$\Delta\phi$ makes an entangled state of the two electrons (one injected
from the upper and the other from the lower interferometers)~\cite{bi-solid}.

Let us consider a two electron injection process, one from lead $\bar{\alpha}$
and the other from $\bar{\gamma}$.
Two types of electron creation operators are introduced, namely
$c_x^\dagger$ and $b_x^\dagger$. The operators $c_x^\dagger$ and $b_x^\dagger$
create an electron at lead $x$ and at the
intermediate regions, respectively.
The beam splitter BS-$i$, made of a quantum point contact, is characterized by
the scattering matrix $S_i$ ($i=1,2,3,4$)
\begin{subequations}
 \label{eq:S-matrix}
\begin{equation}
 S_i = \left(
       \begin{array}{cc}
         r_i & t_i' \\
         t_i & r_i'
       \end{array} \right) \;,
\end{equation}
which transforms the electron operators as
\begin{eqnarray}
 ( \begin{array}{cc}
          c_{\bar{\alpha}}^\dagger & c_{\bar{\beta}}^\dagger
        \end{array} )
 &=&  ( \begin{array}{cc}
          b_\alpha^\dagger & b_\beta^\dagger
        \end{array} ) S_1, \;\;
 ( \begin{array}{cc}
          b_\alpha^\dagger & b_\beta^\dagger
        \end{array} )
 =  ( \begin{array}{cc}
          c_\alpha^\dagger & c_\beta^\dagger
        \end{array} ) S_2, \\
 ( \begin{array}{cc}
          c_{\bar{\gamma}}^\dagger & c_{\bar{\delta}}^\dagger
        \end{array} )
 &=&  ( \begin{array}{cc}
          b_\gamma^\dagger & b_\delta^\dagger
        \end{array} ) S_3, \;\;
 ( \begin{array}{cc}
          b_\gamma^\dagger & b_\delta^\dagger
        \end{array} )
 =  ( \begin{array}{cc}
          c_\gamma^\dagger & c_\delta^\dagger
        \end{array} ) S_4.
\end{eqnarray}
\end{subequations}

The injected two electon state prior to interaction at the intermediate region
is written as
\begin{equation}
 \left|\Psi_0\right\rangle = c_{\bar{\alpha}}^\dagger
     c_{\bar{\gamma}}^\dagger|0\rangle
      = ( r_1b_\alpha^\dagger+t_1b_\beta^\dagger )
     \otimes ( r_3b_\gamma^\dagger+t_3b_\delta^\dagger ) |0\rangle ,
\end{equation}
where $|0\rangle$ is the ground state without injection.
It is assumed that Coulomb interaction affects only the trajectory
of the two electrons and inelastic scattering is neglected. This
assumption can be justified at low bias and temperature. The
modification of trajectory changes the area enclosed by the loops of
the interferometers and induces an additional phase shift
$\Delta\phi$
\begin{equation}
\Delta\phi = 2\pi H\Delta A/\Phi_0 \,,
 \label{eq:Delta-phi}
\end{equation}
for the particular path in which both electrons are transmitted. $H$
and $\Delta A$ stand for the external magnetic field and the area
enclosed by the change of the trajectory resulting from the
interaction (represented by the shaded region of Fig.~1),
respectively. $\Phi_0=hc/e$ is the flux quantum. As a result, the
two-electron state upon a scattering can be written as
\begin{subequations}
 \label{eq:Psi}
\begin{equation}
 \left|\Psi\right\rangle =
   ( r_1 b_\alpha^\dagger \chi_r^\dagger
   + t_1 b_\beta^\dagger \chi_t^\dagger
   ) |0\rangle \;,
\end{equation}
where the operators $\chi_r^\dagger$ and $\chi_t^\dagger$
create the states of the lower interferometer
depending on whether the electron in the upper interferometer
is reflected or transmitted,
respectively:
\begin{eqnarray}
 \chi_r^\dagger &=& r_3 b_\gamma^\dagger + t_3 b_\delta^\dagger \;, \\
 \chi_t^\dagger &=& r_3 b_\gamma^\dagger
                 + t_3 e^{i\Delta\phi} b_\delta^\dagger \;.
\end{eqnarray}
\end{subequations}
Eq.~(\ref{eq:Psi}) describes an entanglement of the two interferometers.
Because of the phase factor $e^{i\Delta\phi}$,
$\chi_r^\dagger\ne\chi_t^\dagger$
and the extent of the entanglement can be controlled by changing $H$ or
$\Delta A$.

The interference of single electrons in the upper interferometer
is displayed in
the probability of finding an electron at lead $a$ ($\in \alpha, \beta$),
\begin{equation}
 P_a = \langle\Psi| c_a^\dagger c_a |\Psi\rangle \;.
\label{eq:Pa}
\end{equation}
The evaluation with the help of
Eqs.~(\ref{eq:S-matrix},\ref{eq:Psi}) gives
\begin{eqnarray}
 P_\alpha &=& 1-P_\beta \label{eq:Pa} \\
   &=& R_1R_2 + T_1T_2 + 2|\nu|\sqrt{R_1T_1R_2T_2}
     \cos{(\varphi-\phi_\nu)} \nonumber\;,
\end{eqnarray}
where $T_i=|t_i|^2$ and $R_i=|r_i|^2$ are the transmission
and the reflection probabilities at BS-$i$, respectively.
The overlap of the states
$\nu\equiv \langle0|\chi_t\chi_r^\dagger|0\rangle$ is a
measure of the WP information
and $\phi_\nu\equiv \arg{\nu}$.
The phase $\phi_1$ enclosed by the loop of the upper interferometer
is given as $\phi_1=\arg{(t_1)}+\arg{(t_2')}-\arg{(r_1)}
-\arg{(r_2)}$.

Eq.~(\ref{eq:Pa}) shows the relation between the interference fringe
of the upper and the WP information stored in the lower
interferometer. If the two states $\chi_r^\dagger|0\rangle$ and
$\chi_t^\dagger|0\rangle$ are orthogonal (that is $\nu=0$), then a
complete WP information is acquired and the interference disappears
in the upper interferometer. Complete WP information can be obtained
for a symmetric BS-3 ($|r_3|=|t_3|=1/\sqrt{2}$) with
$\Delta\phi=\pi$. Note that $\nu$ is unaffected by the scattering at
BS-4.

For studying the nonlocal correlation between the two subsystems
we calculate the joint detection
probabilities $P_{ab}$ of two electrons (one from lead $a\in\alpha$ or $\beta$
and the other from lead $b\in\gamma$ or $\delta$). For instance,
joint-detection probability at leads $\alpha$ and $\gamma$ is given by
\begin{equation}
 \label{eq:Pac}
 P_{\alpha\gamma} = |r_1r_2u_\gamma + t_1t_2'v_\gamma|^2
                                \;,
\end{equation}
where the coefficient $u_\gamma\equiv r_3r_4+t_3t_4'$
($v_\gamma\equiv r_3r_4+t_3t_4'e^{i\Delta\phi}$) represents the
amplitude of finding an electron at lead $\gamma$ under the
condition that the electron in the upper interferometer is reflected
(transmitted) at BS-1.

In ballistic conductors, it is practically impossible to count
electrons one by one. Instead, the single-particle and
joint-detection probabilities can be obtained by measuring the
average current and the zero-frequency cross
correlation~\cite{kang07}.  The output electrodes are grounded with
zero voltage. The input electrodes ($\bar{\alpha},\bar{\gamma}$) are
biased by the voltage $eV$. We assume that the transmission of
entangled electrons can be written down as an `entangled many-body
transport state' of the form
\begin{equation}
\label{eq:Psi_bar}
 |\bar{\Psi}\rangle = \prod_{0<E<eV} \left[
   r_1 b_\alpha^\dagger(E) \chi_r^\dagger(E)
   + t_1 b_\beta^\dagger(E) \chi_t^\dagger(E)
   \right] |\bar{0}\rangle \;,
\end{equation}
where $|\bar{0}\rangle$ stands for the filled Fermi sea in
all leads at energies $E<0$.
The main assumption made here is that the injected electrons
from the two sources interact with each other and are
transmitted as entangled pairs as illustrated in Fig.~1.

For the state $|\bar{\Psi}\rangle$, the output current at lead $a$
($I_a$) is proportional to the probability of finding an electron at
this lead, $I_a=(e^2/h)P_aV$. The zero-frequency cross correlation,
$S_{ab}$, of the current fluctuations $\Delta I_a$ and $\Delta I_b$,
is defined as~\cite{buttiker90}
\begin{equation}
 S_{ab} = \int dt\, \langle\bar{\Psi}|
   \Delta I_a(t)\Delta I_b(0)+\Delta I_b(0)\Delta I_a(t)
        |\bar{\Psi}\rangle \;.
\end{equation}
For $a\in\alpha,\beta$ and $b\in\gamma,\delta$, this cross correlator
provides information about the two-particle interactions between the two
interferometers which is expressed in terms of the following useful relation:
\begin{equation}
 S_{ab} = \frac{2e^2}{h} eV (P_{ab}-P_aP_b) \;.
 \label{eq:Sab}
\end{equation}
Therefore, the joint-detection probabilities can be obtained by measuring
the average current and cross correlation.

BI is a test to distinguish quantum correlation from
local hidden variable theory by considering two-particle
correlation~\cite{bell64}.
Let us first consider the maximally entangled case: BS-1 and BS-3 are
symmetric and $\Delta\phi=\pi$.
General case with arbitrary $\Delta\phi$ is
discussed later.  BS-2 is also being kept symmetric,
and the phase enclosed by the loop of the lower interferometer ($\phi_2$)
is fixed at $\varphi_2=\pi/2$.
The phase of the upper loop
($\phi_1$) and the transmission probability of BS-4 ($T_4$) are controlled
to test the BI. Adopting a new phase variable $\theta$ as
$T_4=\sin^2{(\theta/2)}$, the BI is written in the form~\cite{chsh69}
\begin{equation}
 S = |E(\phi_1,\theta)-E(\phi_1',\theta)+E(\phi_1,\theta')+E(\phi_1',\theta')
     | \leq 2 ,
\label{eq:chsh}
\end{equation}
where
\begin{equation}
E(\phi_1,\theta) = P_{\alpha\gamma}+P_{\beta\delta}-
                   P_{\alpha\delta}-P_{\beta\gamma}
   = \cos{(\phi_1-\theta)} .
\end{equation}
The maximal violation of the BI is found for
$\phi_1=0,\phi_1'=\pi/2,\theta=-\pi/4$, and $\theta'=\pi/4$, where
the maximum Bell parameter is $S_{max}=\max{S}=2\sqrt{2}$.

For an arbitrary value of $\Delta\phi$ (i.e., for an arbitrary degree of
entanglement), we also find the maximum Bell parameter by optimizing
the phases $\phi_1,\phi_2$ and $\theta$. It reads
\begin{equation}
 S_{max} = 2\sqrt{1+\sin^2{(\Delta\phi/2)}} ,
\end{equation}
i.e., the BI is violated for nonzero $\sin{(\Delta\phi/2)}$.
This implies that the BI is violated for any
entangled pure state~\cite{gisin91}. Note that the state of Eq.(\ref{eq:Psi})
cannot be written in a product state of two electrons unless
$\sin{(\Delta\phi/2)}=0$.

In conclusion, Bell's inequality can be tested in coupled electronic
Mach-Zehnder interferometers. Entanglement of electrons is generated via
Coulomb-interaction-induced modification of trajectories. This entanglement
suppresses interference fringe for the
single electron transport. Investigation on the two-electron correlation
shows that the Bell's inequality is violated for any degree of entanglement.

{\em Acknowledgments} -
We thank J. Lee for helpful comments.
This work was supported by the Korea Research Foundation
(KRF-2005-070-C00055,  KRF-2006-331-C00116).

\begin{figure}[h]
 \begin{center}\leavevmode
 \includegraphics[width=0.7\linewidth]{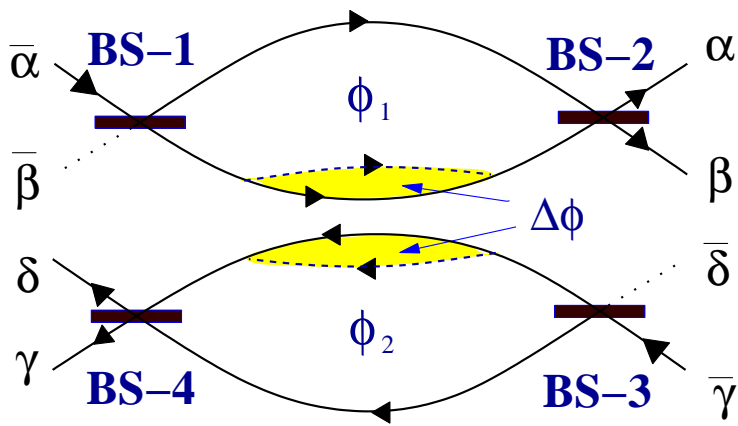}
 \caption{Schematic figure of the coupled electronic Mach-Zehnder interferometers
  to test the Bell's inequality}
 \label{fig1}
 \end{center}
\end{figure}

\end{document}